\documentclass[sigconf,nonacm]{acmart}

\AtBeginDocument{%
  \providecommand\BibTeX{{%
    \normalfont B\kern-0.5em{\scshape i\kern-0.25em b}\kern-0.8em\TeX}}}

\usepackage[linesnumbered,ruled,vlined]{algorithm2e}

\usepackage{graphicx}
\usepackage{booktabs}
\usepackage{multirow}

\acmConference[KDD]{KDD}{2021}{Singapore}

\begin{document}

\renewcommand\footnotetextcopyrightpermission[1]{}
\renewcommand{\thefootnote}{\fnsymbol{footnote}} 

\title{Truncation-Free Matching System for Display Advertising at Alibaba}
\author{Jin Li$^*$, Jie Liu$^*$}
\thanks{$^*$Both authors contributed equally to this work.}
\author{Shangzhou Li, Yao Xu, Ran Cao, Qi Li, Biye Jiang}
\author{Guan Wang, Han Zhu, Kun Gai, Xiaoqiang Zhu}
\affiliation{
  \institution{Alibaba Group}
  \city{Beijing}
  \country{China}
}
\email{{echo.lj,jierui.lj}@alibaba-inc.com}
\email{{shangzhou.lsz,zaifeng.xy,caoran.cr,luyuan.lq,biye.jby}@alibaba-inc.com}
\email{{shangfeng.wg,zhuhan.zh,jingshi.gk,xiaoqiang.zxq}@alibaba-inc.com}

\renewcommand{\shortauthors}{Li et al.}

\newcommand{\LINJI}[1]{{\color{blue} LINJI: #1}}
\newcommand{\JIERUI}[1]{{\color{green} JIERUI: #1}}
\newcommand{\HUAIREN}[1]{{\color{red} HUAIREN: #1}}
\newcommand{\DASHI}[1]{{\color{purple} DASHI: #1}}
\newcommand{\YIFANG}[1]{{\color{yellow} YIFANG: #1}}
\newcommand{\argTopn}{\operatornamewithlimits{argTop-n}}
\newcommand{\argTopm}{\operatornamewithlimits{argTop-m}}
\newcommand{\argTopk}{\operatornamewithlimits{argTop-k}}
\newcommand{\Topn}{\operatornamewithlimits{Top-n}}
\newcommand{\Topm}{\operatornamewithlimits{Top-m}}
\newcommand{\Topk}{\operatornamewithlimits{Top-k}}

\begin{abstract}

 Matching module plays a critical role in display advertising systems. Different from sponsored search where user intentions can be captured naturally through query, display advertising has no explicit information about user intentions. Thus, it is challenging for display advertising systems to match user traffic and ads suitably w.r.t. both user experience and advertising performance.  \textbf{From the advertiser's view}, system packs up a group of users with common properties, such as the same gender or similar shopping interests, into a \emph{crowd}. Here term \emph{crowd} can be viewed as a tag over users in the same crowd. Then advertisers bid for different crowds and deliver their ads to those targeted users. \textbf{From the advertising system's view}, things turn to be a little different. So far as we know, matching module in most industrial display advertising systems follows a two-stage paradigm. When receiving a user visit request,  matching system (i) finds the \emph{crowds} that the user belongs to; (ii) retrieves all ads that have targeted those crowds. However, in real world applications, such as the display advertising at Alibaba, with volume of crowds reaching up to tens of millions and volume of ads reaching up to millions, both stages of matching have to truncate the long-tailed user-crowd or crowd-ad pairs for online serving, under limited latency and computation cost requirements. That is to say, not all advertisers that bid for a given user have the chance to participate in the online matching process. This results in sub-optimal advertising performance for advertisers. Besides, it also brings loss of revenue of the advertising platform.  
 
 In this paper, we study carefully the truncation problem and propose a \textbf{T}runcation-\textbf{F}ree \textbf{M}atching \textbf{S}ystem (TFMS). The basic idea of TFMS is to decouple the matching computation from the online processing pipeline. Instead of executing the two-stage matching when user visits, TFMS utilizes a near-line truncation-free matching module to pre-calculate and store those top valuable ads for each user. Then, the online pipeline just needs to fetch the pre-stored candidate ads as the result of matching. In this way, near-line matching can jump out of the online system's latency and computation cost limitations and leverage flexible computation resources to finish the user-ad matching process. Moreover, we can employ arbitrary advanced models to conduct the top-n candidate selection in the near-line matching system over all candidate ad set, bringing superior performance compared with original roughly truncated online matching system. Since 2019, TFMS has been deployed in our productive display advertising system, bringing (i) more than $50\%$ improvement of impressions for advertisers who encountered truncation before,  (ii) $9.4\%$ RPM (Revenue Per Mile) gain for advertising system, which is significant enough for the business.

\end{abstract}



\maketitle

\section{Introduction}

Online display advertising has played an important role in marketplace in last several decades \cite{evans2009online,goldfarb2011online}. Advertisers bid for online traffic according to their marketing demands. For example, a clothing shop bids for users who have shopping needs of clothes through the advertising system when these users visit Alibaba's e-commerce site. Different from search advertising (also known as sponsored search) where user intentions can be captured naturally through query, display advertising has no explicit information about user intentions. Thus, it is challenging for display advertising systems to match user traffic and ads suitably w.r.t. both user experience and advertising efficacy. 

Practically, display advertising systems usually provide tools for advertisers to select the target audiences.  This is called \textbf{targeting} in online advertising term. Taking the display advertising system at Alibaba as an example, there are different kinds of targeting ways for advertisers to select audiences they want to bid for. We list several typical ones as follows:
\begin{itemize}
	\item \emph{Retargeting}: to target users that have already viewed, clicked or even added to cart the advertiser's products (or similar products) recently.   
	\item \emph{Keywords targeting}: to target users who have interests that are expressed by the keywords, such as `travel', `swim', etc.      
	\item \emph{Demographic targeting}: to target users according to their demographic information, such as age, gender, etc. 
	\item \emph{Automatic targeting}: to target users automatically by the system, which searches suitable users for advertisers with advanced models learned from large-scale historical user behavior data. For example, TDM \cite{zhu2018learning, zhu2019joint, zhuo2020learning} system has served as an important automatic targeting tool for the display advertising system at Alibaba. 
\end{itemize}

With the help of these targeting tools, now advertisers can select target audiences and bid for them with differentiated bidding prices. 
When a user visits the e-commerce site, the display advertising system receives a traffic request. Then system processes the request in three steps usually: (i) \textbf{Matching Step}: checks all the involved ads that bid for this user and selects top-$n$ suitable candidate ads. Usually active users can be bid by tens of thousands of ads. In this way, only a part of ads can be chosen to send to the next step, considering the ads' revenue and relevance in a rough manner. (ii) \textbf{Ranking Step}: evaluates precisely each ads' revenue with advanced models \cite{zhou2018deep,zhou2019deep,wang2018dkn} and ranks them. (iii) \textbf{Auction Step}: decides which ads win the auction and calculates the cost respectively \cite{zhu2017optimized,jin2018real}. It is worth mentioning that matching step mainly has two aspects of functionalities: 
\begin{itemize}
	\item Matching system should ensure that advertisers can reach each target audiences they bid for; 
	\item Matching system should be responsible for the advertising platform's revenue and user experience.
\end{itemize}
 
\begin{figure}[h]
  \centering
  \includegraphics[width=0.8\linewidth]{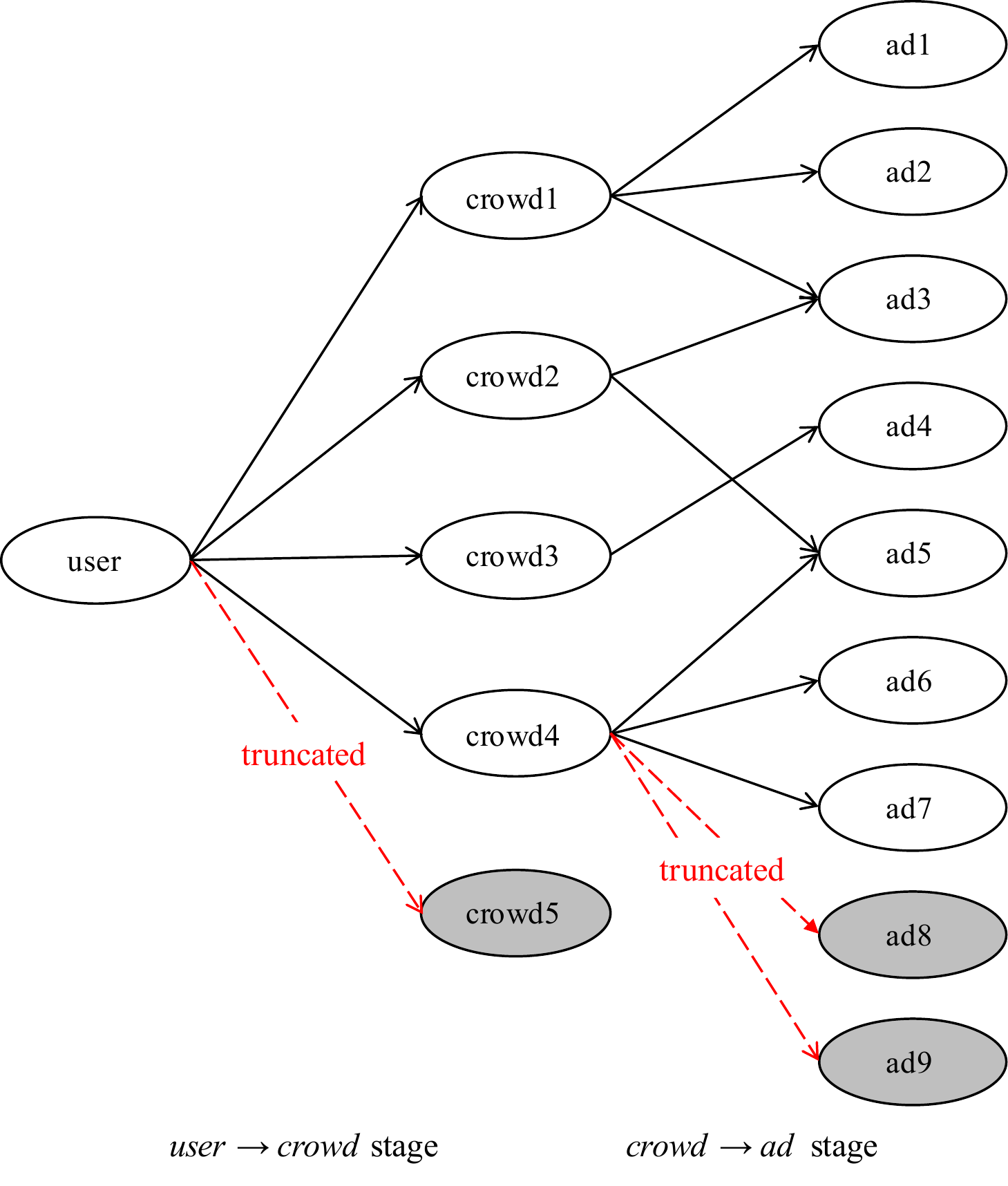}
  \caption{Illustration of the two-stage matching process in the display advertising system. Candidate ads are selected by execution of `user -> crowd' and `crowd -> ad' stages. Truncation strategies are used in both stages to reduce long-tail calculation for saving online latency and computation cost.}
  \label{fig:truncated}
\end{figure}

So far as we know, matching module of most industrial display advertising systems is designed under a two-stage paradigm. When receiving a traffic request, a.k.a. a user visit, the first stage of matching is to find the \textbf{crowds} that the user belongs to. The term `crowd' in display advertising system is a concept associated with advertiser's targeting. `crowd' can be viewed as tag information that users in the same crowd own. It is the bridge to link users and ads. For example, if an ad campaign decides to bid for users whose age ranging from 18 to 25 through demographic targeting, then all users with age ranging from 18 to 25 is a crowd. After finding the crowds, the second stage of matching is to retrieve all the ads that have targeted those crowds. So far, the matching stage successfully find valid ad candidates. Figure~\ref{fig:truncated} illustrates the two-stage matching process. 

Actually, the two-stage matching architecture exists widely in industrial applications apart from display advertising system. For example, when handling user request in search engine, a (\emph{user query} -> \emph{keywords} -> \emph{documents}) two-stage retrieval process is executed based on inverted index \cite{cutting1989optimization}. In recommender systems, the item-based collaborative filtering algorithm follows the (\emph{user} -> \emph{trigger item} -> \emph{similar item}) pipeline. Without loss of generality, we can summarize this kind of two-stage matching architecture into a same pattern, i.e., (\emph{user} -> \emph{intermediate token} -> \emph{item}).

For display advertising systems in a start-up business, both the volumes of `crowds' and `ads' are small enough and the two-stage matching system can run perfectly. However, in most industrial cases, such as the display advertising at Alibaba, with volume of crowds reaching up to tens of millions and volume of ads reaching up to millions, the \textbf{truncation problem} of the two-stage matching system will bring severe hurt. Truncation exists in both (\emph{user} -> \emph{crowd}) and (\emph{crowd} -> \emph{ad}) stages of the matching system, as illustrated in Figure~\ref{fig:truncated}. In few words, to finish the matching step under limited latency and computation cost, real world display advertising systems often truncate (or even skip) those long-tail user-crowd pairs and crowd-ad pairs. Thus, not all advertisers that bid for a given user have the chance to participate in the matching process. Moreover, the truncation strategy is often determined in an offline manner, which is usually executed when building the \emph{user} -> \emph{crowd} index and \emph{crowd} -> \emph{ad} index. It means that advertisers be truncated will never have the chance to participate in the matching process unless the offline built indices are re-built or updated. In our system, the volume of ads that linked to a given crowd cannot exceed 2,000, which is carefully tuned to meet system latency and computation cost requirements. Table~\ref{tb:trunc-stat} shows the truncation statistics in our system. Obviously, truncation results in sub-optimal advertising performance for advertisers. Besides, it also brings loss of revenue to the advertising platform.  

\begin{table}[htbp]
\caption{Truncation statistics in the display advertising system at Alibaba}\label{tb:trunc-stat}
\begin{tabular}{lll}
	\toprule
    & User-Crowd  & Crowd-Ad  \\\midrule
  Percentage of truncated pairs   & 36\% & 21\% \\
  \bottomrule
\end{tabular}
\end{table}

In this paper, we rethink the challenge from a system design's view and propose a truncation-free solution which has already been deployed in the display advertising system at Alibaba. We observe that the key conflict lies between the system limitation (limitation of online latency and the computation cost) and the long-tail volumes of user-crowd and crowd-ad pairs. Based on this insight, we propose a new architecture for the matching system towards truncation-free matching. The basic idea is to decouple the matching computation from the online processing pipeline. Instead of executing the two-stage matching when user visits, our new design utilizes a near-line truncation-free matching module to pre-calculate and store those top valuable ads for each user. Then, the online pipeline just needs to fetch the pre-stored candidate ads as the result of matching. In this way, near-line matching computation can jump out of the limitation of online system and leverage flexible computation resources to finish the user-ad matching process. Moreover, we can employ arbitrary advanced models to conduct the top-n candidate selection in the near-line matching system, bringing superior performance compared with original roughly rule-based truncation in online matching system. We name the proposed solution as \textbf{T}runcation-\textbf{F}ree \textbf{M}atching \textbf{S}ystem (TFMS). 

To the best of our knowledge, our paper is the first work that studies carefully the widely existed truncation problem in matching system of display advertising and propose a truncation-free solution. Since 2019, TFMS has been deployed in our productive display advertising system. Advertisers that encountered truncation in original two-stage online matching system gets more than $50\%$ improvement of impressions in TFMS. Besides, advertising platform also enjoys benefits. For example, in our \emph{banner ads} product, TFMS brings $9.4\%$ RPM (Revenue Per Mile) gain, which is significant enough for the business. It's worth mentioning that the proposed truncation-free near-line matching solution not only works well in the display advertising system, but also is a general framework for other applications that use truncated two-stage matching. We hope that our work could bring some motivations for these applications from a new perspective. 

The rest of this paper is organized as follows: Section~\ref{sec:related} introduces some background of crowd targeting in display advertising. Section~\ref{sec:design} and \ref{sec:system} describe in detail the design of the proposed TFMS. Section~\ref{sec:implement} describes how we implement TFMS in our productive advertising system. Section~\ref{sec:exp} gives experiments about TFMS compared with traditional two-stage online matching system. Section~\ref{sec:conclusion} concludes our work.

\section{Background}\label{sec:related}

In this section, we give a brief introduction of matching module in the display advertising of Alibaba, especially explains in detail the truncation problem, to help readers understand why we need to design a novel truncation-free matching system. 

\subsection{Truncated Two-stage Matching Process in Display Advertising System of Alibaba}
Like many other industrial systems, matching module in the display advertising system of Alibaba follows a common (\emph{user -> crowd -> ad}) two-stage structure to retrieve candidate ad set. Figure~\ref{fig:truncated} illustrated this process. 
\begin{itemize}
	\item \textbf{(\emph{user} -> \emph{crowd}) stage}. As described in the introduction section, we provide different targeting tools for advertisers for audience selection. We name each tool as a \emph{matching channel} in the online serving manner. That is, for each user, a multi-channel way is used to find the crowds he belongs to.  Each channel handles a specific type of crowds, like retargeting channel, keywords targeting channel, or model-based automatic targeting channel. Crowds collected from each channel are then merged together. 
	\item \textbf{(\emph{crowd} -> \emph{ad}) stage}. Given crowds merged from each channel, ads that have targeted those crowds are retrieved from an inverted index which contains \emph{crowd} -> \emph{ad} pairs. Considering the computing cost of online serving, a pre-ranking module \cite{wang2020cold} is also applied to further reduce the size of user-ad pairs for the following ranking module. However, due to system performance limitation, here the pre-ranking model has to been simple enough. 
\end{itemize}

\subsection{Truncation Problem}
Facing with billions of traffic every day, truncation strategies are widely used in both stages to meet the latency and computation cost requirements for online serving. 

In \emph{user} -> \emph{crowd} stage, an active user usually belongs to thousands of effective crowds, even in a single channel. Here \emph{effective} means these crowds are targeted by advertisers. For example, a young mother may have diverse interests including  milk powder shopping, yoga fitness, swimming, traveling etc. She may be targeted by advertises from all these businesses.  If all these crowds are sent to the second stage, it will cause serious latency problem in processing real-world traffic. Practically, we truncate the crowd volume for each channel separately, considering the  \emph{crowd} -> \emph{ad} volume for different type of  crowds. For example, a crowd of \emph{20 to 25 years old young girls} from demographic targeting may link tens of thousands of ads, while a crowd of \emph{people who have added my product to cart} might link only dozens of ads. In our real system, the truncation threshold for crowd volume is usually less than 100. Here truncation is conducted in each channel by its own truncation model, which can be viewed as some rule-based statistical scores.  Obviously, it is rough and has no connection with ads that have targeted the truncated crowds.  

In \emph{crowd} -> \emph{ad} stage, as described above, a popular crowd  may be targeted by tens of thousands of advertisers. To retrieve all these long-tail candidate ads from this crowd is not acceptable, in which latency cost will exceed the limitation for online serving.  To overcome this, a limitation of ads number for a single crowd is employed by performing an offline pre-truncation strategy. Again, truncation is conducted by some rule-based approaches, such as average CTR of each ad in the last 7 days etc.  In our system, the volume of ads that linked to a given crowd cannot exceed more than 2000, which is carefully tuned to meet the latency and computation cost requirement.

\begin{table}[th]
\caption{Simulated performance of two stages by applying truncation-free strategy directly in our real system}\label{tb:offline-overview}
\begin{tabular}{lll}
	\toprule
  Truncation-Free Stage  & Latency & Simulated Revenue \\\midrule
  (\emph{user} -> \emph{crowd}) & +26\% & +5.5\% \\
  (\emph{crowd} -> \emph{ad})  & +17\% & +2.5\% \\
  both stages & +51\% & +5.6\% \\
  \bottomrule
\end{tabular}
\end{table}

To further get an intuitive understanding,  we make online simulations in our real system by deploying truncation-free strategy and without considering the latency and computation cost. These simulations are performed using logged data. Table~\ref{tb:offline-overview} gives the results. Here we only calculate the impact from advertising platform's view. Obviously, without truncation the two-stage matching is of higher revenue theoretically but unacceptable latency, as in our real system an increase of $10\%$ latency is unacceptable. 

Now it is clear to see the hurt from truncation problem.  Here we summarize it formally: 
\begin{itemize}
	\item Truncation causes those advertisers that have bid for selected target audiences to lose chance for participating in the matching process. This is unfair for them, as even these advertisers further rise the bid price, things will be the same.  
	\item Truncation causes the advertising platform to lose chance to evaluate precisely each candidate ad, thus resulting in sub-optimal platform revenue and user experience.  
\end{itemize}



%


\section{Matching Problem in Display Advertising}\label{sec:design}
To tackle the challenges caused by truncation in the above discussed two-stage online matching system, in this section we give detailed review of the matching problem in display advertising, thus derive a truncation-free solution. It is worth mentioning that approaches discussed here can also be extended to other applications using the same two-stage matching architecture.

\subsection{Definition of Matching Problem in Display Advertising}

Given a user $u$, let $\mathbf C(u)$ be the set of all valid crowds that the user belongs to:
\begin{equation}
	\mathbf C(u) = \{c~|~u ~\text{belongs to}~ c\}.
\end{equation}
Here $c$ denotes a crowd. Given $c$, let $\mathbf A(c)$ be the set of all valid ads that have selected $c$ as the targeting crowd:
\begin{equation}
	\mathbf A(c) = \{ a~|~ \text{crowd} ~ c ~ \text{is targeted by} ~ a \}.
\end{equation}
Then we can define the total valid candidate set $\mathbf O(u)$  for user $u$ as
\begin{equation}
	\mathbf O(u) = \{ (a, c)~| c \in \mathbf C(u) \wedge a \in \mathbf A(c) \}.
\end{equation}
We treat each element in $\mathbf O(u)$ as an ad-crowd pair as that advertiser may have different bid prices on different target crowds even for the same ad, i.e., ${bid}(a, c_1) \neq {bid}(a, c_2)$.

Now, the objective of matching in display advertising can be viewed as an optimization problem w.r.t. a matching system  $f$:
\begin{equation}\label{eq:problem}
	\begin{aligned}
		&{\underset {f} {\operatorname {maximize} }} && \sum_{u} R(f(u))  \\
		&\operatorname {subject\ to} && f(u) \subset \mathbf O(u) \\
		&&& L(f(u)) \leq l \\
		&&& card(f(u)) = n,
	\end{aligned}
\end{equation}
where $f(u)$ is the matching result of user $u$ given matching system $f$, $R(\cdot)$ is a metric function to measure the reward of both advertiser's performance (as well as fairness) and advertising platform's revenue, $L(\cdot)$ measures the system performance of $f$, such as latency, computation cost, etc. $card(\cdot)$ measures the volume of matching result $f(u)$. The goal of designing a matching system $f$ is to select a set of candidate ads $f(u)$ with size $n$ from all valid candidates $\mathbf O(u)$ to maximize the reward $R$, while satisfying the requirement of online system performance $l$. It is trivial that $\sum_{u} R(f(u))$ reaches the maximal value when $f(u)$ equals to $\mathbf O(u)$. That is, matching system $f$ returns all the valid $(a, c)$ pairs. In this case, both the advertisers and the advertising platform get the maximal satisfaction.
  
Practically, it is not straightforward to measure the satisfaction from advertiser side in reward function $R(\cdot)$. But it is easy to prove that when advertising platform's revenue reaches maximal value, advertisers' performance also reaches maximal point \cite{krishna2009auction}. Thus, without loss of generality, in the rest of this paper, we simplify reward function $R(\cdot)$ to be calculated just from the advertising platform side. 

Given user $u$, the reward of $R(f(u))$ can be substituted by the averaged value across set $f(u)$:
\begin{equation}	
	R(f(u)) = \frac{\sum_{(a_i, c_i) \in f(u)} r_u(a_i, c_i)}{n}, 
\end{equation}
where $r_u$ is some value measurement (e.g., eCPM in display advertising system) for a given ad-crowd pair $(a, c)$ for user $u$. 
 
\textbf{Optimal matching system.} Based on the above definition, if we ignore the system performance requirement, i.e., $L(f(u)) \leq l$, the optimal solution of matching system $f$ would be to select the set of top-$n$ $(a, c)$ pairs from $\mathbf O(u)$ under $r_u$ measure:
\begin{equation}
\label{equ:opt}
	f_{opt}(u) = \argTopn_{(a, c) \in \mathbf O(u)} r_u(a, c).
\end{equation}
Note that here \emph{optimal solution} should also takes into consideration of advertisers' dynamic operation during the whole advertising period, since that $r_u$ is a real-time value measurement.



In real-world business, without truncation, the volumes of valid ad set $\mathbf O(u)$ are usually larger than the ones that can be afforded by the online serving system. As introduced in Table ~\ref{tb:offline}, the full set of $\mathbf O(u)$ is $259\%$ larger than the truncated one on average. Thus, traditional truncated two-phase matching usually uses truncation strategies to shrink $O(u)$ to a subset of proper size. We conclude and denote the solution of truncated two-stage online matching system as $f_1$:

\begin{equation}\label{eq:twostage}
	\begin{aligned}
	\mathbf C_1(u) &=&& \argTopm_{c \in \mathbf C(u)} ~ s_u(c), ~ \mathbf C_1(u) \subset \mathbf C(u)\\
	\mathbf A_1(c) &=&& \argTopk_{a \in \mathbf A(c)} ~ s_u(a), ~ \mathbf A_1(c) \subset \mathbf A(c)\\
	\mathbf O_1(u) &=&& \{(a, c)~|~c \in \mathbf C_1(u) \wedge a\in \mathbf A_1(c) \}, ~ \mathbf O_1(u) \subset \mathbf O(u)\\
	f_1(u) &=&& \argTopn_{(a, c) \in \mathbf O_1(u)} r_u(a, c) \\
	\end{aligned}
\end{equation}
where $s_u(c), s_u(a)$ denote score functions for crowd $c$ and ad $a$ respectively, i.e., rule-based statistics in our real system; $m, k$ are the truncation numbers for (\emph{user} -> \emph{crowd}) and (\emph{crowd} -> \emph{ad}) stages. Not surprisingly, we have:
\begin{equation}
	R(f_{1}(u)) < R(f_{opt}(u)),
\end{equation}
that is, truncated two-phase online matching system yields a non-optimal solution. Besides, for $(a, c)$ pairs truncated by $f_1$, i.e., the pairs in set $\mathbf O(u) \backslash \mathbf O_1(u)$, they lose the chance to bid for user $u$. This arises unfairness and brings bad bidding experiences for advertisers, which is also harmful for the advertising platform's both short term and long-term revenue.   

\begin{figure*}[htbp]
  \centering
  \includegraphics[width=0.7\textwidth]{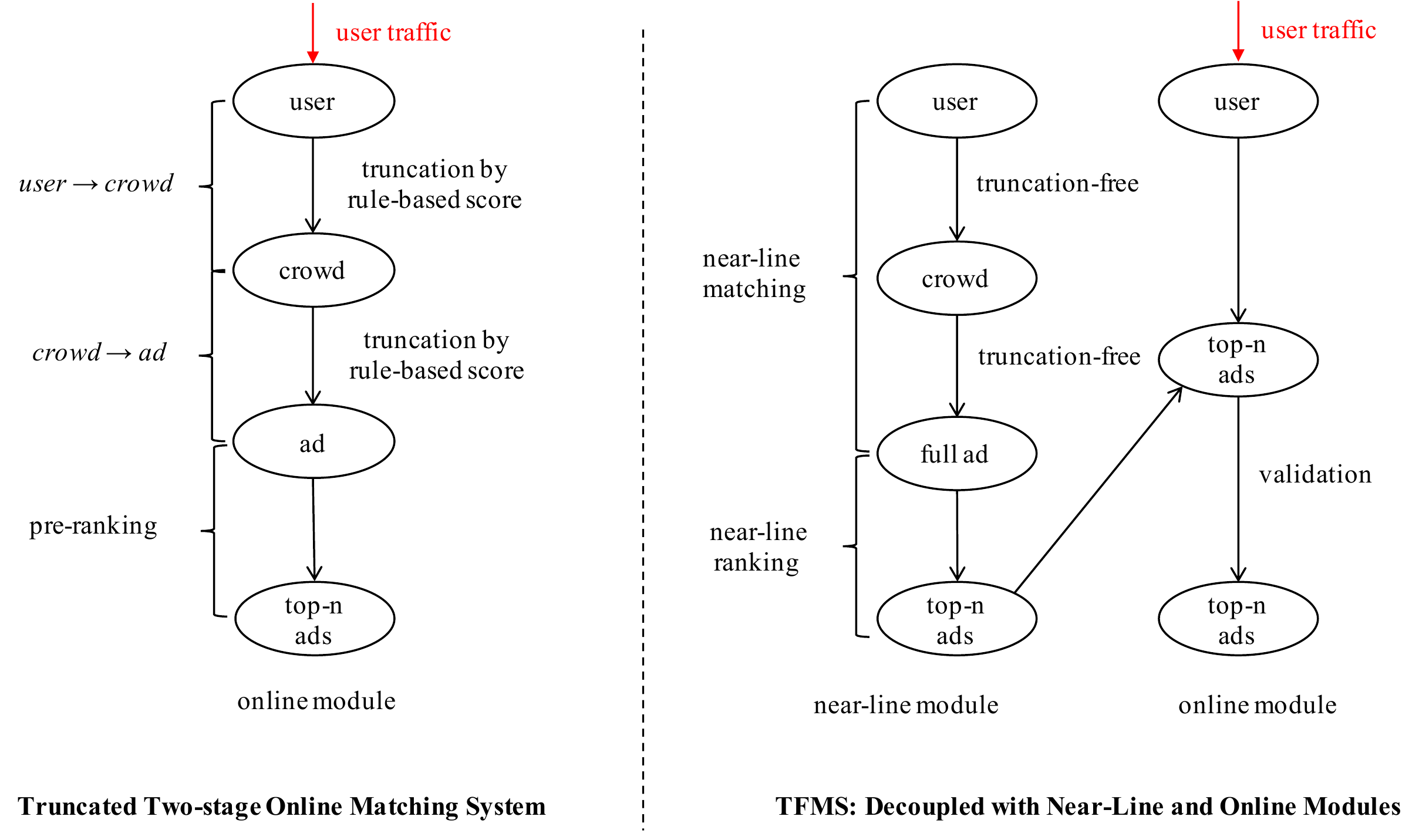}
  \caption{Comparison between traditional two-stage online matching system and our proposed TFMS. Different from two-phase online matching process, which is performed only in online system, TFMS decouples matching process into near-linear calculation and online fetch parts, which enables to achieve truncation-free matching.}
  \label{fig:generation}
\end{figure*}

\subsection{TFMS: a design of optimal matching system}
The brute-force calculation over all valid candidate ads in the online serving manner is of great challenge, under both latency and computation cost limitation. In the past years, we have tried our best to optimize the system performance to allow more ads into the online serving system. However, with the advance of our advertising business, more advertisers enter our system, making the truncation problem more serious. 

Rethinking the problem, we find the truncation is actually caused by the long-tail distribution of \emph{user} -> \emph{crowd} pairs or \emph{crowd} -> \emph{ad} pairs. To cover these long-tail calculations fully in an online serving manner seems very difficult, especially under a strict latency limitation. On the other hand, the matching process is repetitively executed  for a given user with multiple visits to the advertising system, that is, the computing cost is wasted in the fully online serving manner. For example, in our system each user visits $8.2$ times on average every day. This motivates us to design a new matching system, which we name as \textbf Truncation-\textbf{F}ree \textbf Matching \textbf System (TFMS).     

\textbf{Near-line matching design.} The key idea of TFMS is to decouples the matching process from online serving and moves it into a near-line system. We use an additional near-line matching module to generate candidate top $(a, c)$ pairs for each user asynchronously, and the online process just needs to fetch the generated candidates. Since the asynchronous near-line calculation has no latency limit, a truncation-free traverse over $\mathbf O(u)$ is feasible. Besides, considering: (i) the full set of $\mathbf O(u)$ is $259\%$ larger than the truncated one on average; (ii) we can save $720\%$ computation with decoupled manner, as each user we need to execute the matching process once; (iii) more flexible resources can be used in the near-line manner, e.g., utility of servers is often low when the traffic is low, such as the early morning, it is affordable for TFMS to fully evaluate full set of $\mathbf O(u)$ and select top-n valuable ads. That is, TFMS is a reasonable design of optimal matching system.  We conclude TFMS by the following matching system $f_2$:
\begin{equation}
	\label{equ:tfms}
	f_2(u) = \text{OnlineFetch} \left[ \text{Near-Line} \left(\argTopn_{(a, c) \in \mathbf O(u)} r_u(a, c)\right) \right].
\end{equation}

\section{System Design of TFMS}\label{sec:system}
In this section, we discuss in detail the design of our proposed TFMS solution. 

\subsection{Comparison between TFMS and truncated two-stage online matching system}

Unlike traditional truncated two-stage matching system that most computation is carried out in online serving manner with real-time user traffic, TFMS adopts an asynchronous near-line solution, as illustrated in Figure~\ref{fig:generation}. TFMS also applies a two-stage process to get the valid candidate ad set, i.e., the (\emph{user} -> \emph{crowd} -> \emph{ad}) process. Both of these two stages are truncation-free with no latency limitation. Note that in TFMS, arbitrary advanced models can be used to evaluate each candidate ad, which greatly enhance the ability of top-n selection. Then, the generated top-$n$ $(a, c)$ pairs for each user is cached for online serving when user visits.  It is worth mentioning that due to the time difference between online fetching and near-line caching, some the of cached top-$n$ $(a, c)$ pairs may no longer be valid. For example, the ad campaign's budget may run out, or the campaign may even be canceled. Thus, a validation process in TFMS's online module is used to filter those invalid cached results.

\subsection{Core design of TFMS}

\begin{figure*}[h]
  \centering
  \includegraphics[width=0.7\textwidth]{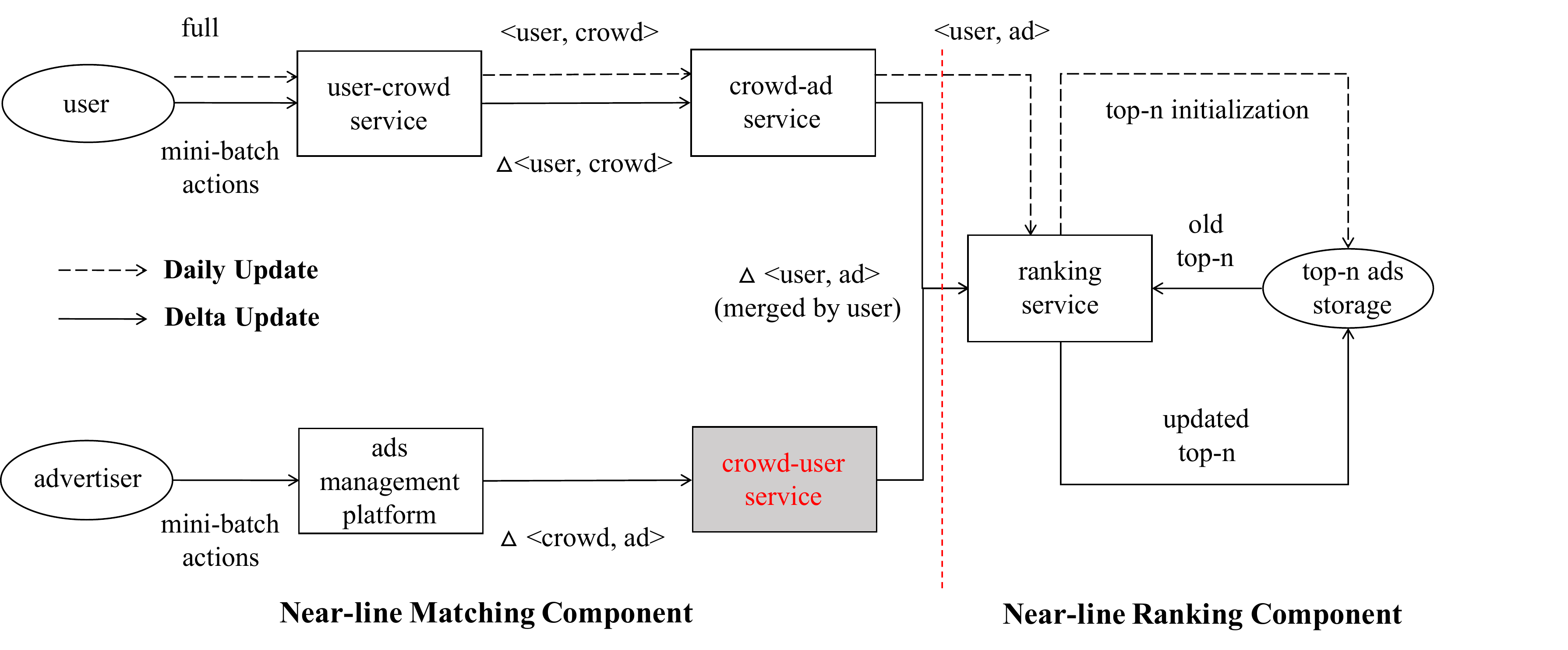}
  \caption{The updating process for top-$n$ ads. Our framework contains a fully update which calculates top-n ads snapshot for all users and a delta update process to update top-$n$ ads based on old top-$n$ ads and delta ads.}
  \label{fig:update}
\end{figure*}

The core part of TFMS is the maintenance of top-$n$ valuable candidate ads for each user for any time.  More specifically, we hope that $f_2$ in Eq.(\ref{equ:tfms}) can be as close to $f_{opt}$ in Eq.(\ref{equ:opt}) as possible. It is clear that keeping the consistency of $\mathbf O(u)$ and $r_u(a,c)$ in online and near-line system of TFMS is the critical point.   

For $\mathbf O(u)$, there are many factors that can arise near-line and online inconsistency. For example, user's newly updated behavior may change the crowds that the user belongs to; advertiser's operation on ad campaigns can influence both (\emph{user} -> \emph{crowd}) and (\emph{crowd} -> \emph{ad}) results. For measure function $r_u(a,c)$, how to keep the consistency depends on the specific value measurement in different advertising businesses. Take the measure function with eCPM (effective cost per mile) in CPC (cost per click) advertising system as an example, prediction value of CTR and ad's bidding price are key factors to keep the consistency of $r_u(a,c)$.  

Based on these observations, TFMS designs a framework that has two update pipelines for keep consistency and maintains the top-$n$ valuable candidate ads: (i) fully update pipeline which only runs once every day as an initialization; (ii) delta update pipeline which runs every 5 minutes incrementally, as illustrated in Figure~\ref{fig:update}.  
\begin{itemize}
	\item \textbf{Fully Update.} We use daily full initialization as the base results of $f_2(u)$. To perform the full initialization, we acquire the truncation-free crowd set $\mathbf C(u)$ for all users, then match the truncation-free ad set $\mathbf A(c)$ for all crowd $c$ in $\mathbf C(u)$ to get $\mathbf O(u)$. Then user-wise top-$n$ valuable ads are ranked and selected by measure function $r_u(a,c)$ for all users.   
	\item \textbf{Delta Update.} Delta update is necessary since that the daily initialization is not able to provide top valuable ads with strong timeliness. Hence, a delta updating mechanism is designed to receive both user and advertiser's actions as input to update the user-wise top ads list. It's worth mentioning that the delta update process triggered by both user and advertiser's action only considers the incremental part of $(a, c)$ pairs that occurred in every delta time interval, such as 5 minutes in our system. Section~\ref{sec:complexity} will give some implementation details about the design of delta update in practice. 
\end{itemize}

Note that, theoretically only the version of TFMS with real-time delta update is a kind of optimal matching system. However, according to our practice, a 5-minutes delta update is approximately optimal, which in turn saves much computation cost since it increases the resource utility of  near-line system in mini-batch manner while not in single request manner. 

To support the two above kinds of update, TFMS further designs two key components, near-line matching and near-line ranking. 

\subsection{Near-line Matching Component}
\label{sec:matching}

Near-line matching component plays an important role in TFMS to provide ad candidates in both full daily initialization and delta update of top valuable ads list. Like traditional truncated two-stage online matching system, the near-line matching component contains a \emph{user-crowd} service and a \emph{crowd-ad} service to conduct (\emph{user} -> \emph{crowd}) and (\emph{crowd} -> \emph{ad}) retrieval to get $\mathbf O(u)$. As illustrated in Figure~\ref{fig:update}, the ads management platform that exists in both traditional online matching and TFMS's near-line matching module handles operations from advertisers to ensure the accurate results of $\mathbf A(c)$ in real time. The near-line matching component mainly has two differences compared with traditional online matching module.

Firstly, benefiting from the near-line design in which exists no latency limitation, the \emph{user-crowd} service and \emph{crowd-ad} service both provide truncation-free retrieval operation. Secondly, an additional \emph{crowd-user} service is added in TFMS's matching component to support user-wise delta update. The \emph{crowd-user} service can be viewed as an inverted version of \emph{user-crowd} service, which can provide (\emph{crowd} -> \emph{user}) look-up operation.  Imagining that when an advertiser changes the targeting crowds of its ad, TFMS's delta update mechanism ensures that each user belongs to the involved targeted crowds be aware of the advertiser's operation and further update those corresponding users' top-n ads list. In this process, the crowd-user service is necessary to help quickly locate the relevant users.

\subsection{Near-line Ranking Component}
\label{sec:preranking}

With the candidate ads generated by near-line matching component, a ranking service including click-through rate or conversion rate prediction is designed to select the final top-n candidate ads. Again, without latency limitation, the near-line ranking component can employ arbitrary advanced models \cite{10.1145/3340531.3412744} to conduct the top-n candidate selection. This ranking process is far more accurate than that in the truncated two-stage online matching system, which consists of ruled-based truncation models and a simple pre-ranking model under strict online serving performance. This obviously brings superior performance for TFMS.

\section{Practice on Implementation of TFMS}
\label{sec:implement}
In section \ref{sec:system} we introduce in detail the ideal design of TFMS. However, in real world applications, we may still suffer from several implementation challenges. Here we share our practice on the implementation of TFMS in our productive display advertising system.

\subsection{Challenge with Huge Volume of Ad Set}
As mentioned earlier, our display advertising system provides different targeting tools for advertisers. Among these targeting tools, automatic targeting is one of the most popular way used by advertisers. In automatic targeting, our platform searches with advanced model to get high quality users for advertisers, which means these ads can be valid candidates for every user. If we consider automatic targeting in TFMS, size of $\mathbf O(u)$ may reach the level of full ad set, i.e. millions in our system, as almost all advertisers have adopted automatic targeting tools. On the other hand, automatic targeting can directly generate top-n candidate ads from the whole ad corpus for each advertiser, without using the two-stage matching process. In our system, full-corpus retrieval models such as ANN\cite{covington2016deep} and TDM\cite{zhu2018learning,zhu2019joint,zhuo2020learning} are developed along another direction, which suffers no truncation problem. Thus, we only implement TFMS with all other targeting tools except automatic targeting.

\subsection{Challenge of Storage}
For each user, top-$n$ valuable candidate ads $f_2(u)$  needs to be stored. In our practice, $n$ cannot be too small. We set it to be 5000 in our real system, considering both resource cost and business performance.  Besides, we also need to maintain the user-crowd storage $\mathbf C(u)$ and the crowd-ad storage $\mathbf A(c)$.  Further, we need to fetch and update these storages very often. Hence, these storages should be both write- and read-friendly. In practice, $\mathbf C(u)$ and $f_2(u)$ are implemented using memory storage like key-value table~\cite{tair2017} and $\mathbf A(c)$ is implemented by inverted index. It is worth to mention that $\mathbf C(u)$ storage is used in both crowd-user and user-crowd service with different key-value setting using the same source data.  
 

\subsection{Challenge of Fully Update}
There are hundreds of millions of users visiting our system every day, making the fully update for all users remaining a great challenge. Clearly, it's impossible for us to calculate $f_2(u)$ for all users at the same time, as it is unfriendly and risky for TFMS which will bring impulse response  to the system.  Besides, not all monthly active users visit our system  every day, which means computation for other in-active users is wasteful. To solve this, we implement the fully update in the following ways: (i) firstly, we only calculate users that have actions in recent days to reduce the computation cost, which can still covers 98\% of daily requests in our system; (ii) secondly, we process active users as streams to share the same stream process structure for updating with parallel strategy to generate $f_2(u)$. That is, the fully update is executed into several parallel streaming pipelines. The parallelism can be adjusted flexibly according to the resources that TFMS can use.   

\subsection{Challenge of Delta Update}\label{sec:complexity}
For delta update, the incremental part caused by advertisers' actions suffers a huge amplification effect, as the volume of crowd can reach tens of millions, which means a single update action of an ad can result in millions of update message. This amplification effect is very unfriendly. Also, the computation cost needed for this kind of update is also huge for us. 

Two types of solutions are considered by us to solve this challenge. One way to handle this kind of pressure is to take a trade-off between real-time performance and computing resources. We use a window function mechanism to aggregate updates for same user in 5-10 minutes to significantly decrease the QPS (query per second) of update process.
Another approximate way of handling this is to discard certain advertiser actions from user view. Only active user will be affected by advertiser action. For users who have actions today, we follow the full update process to recompute $f_2(u)$ after user's action. In this way, advertiser's action that affect this active user of today will be renewed automatically. In practice, we implement the first strategy. 


\section{Experiments}\label{sec:exp}

In this section, we give some offline and online results about our investigation and practice on TFMS for display advertising at Alibaba. Through the experimental results, we hope to answer three questions:

{\bf RQ1: } Whether truncation is a severe problem in traditional two-stage matching or not?

{\bf RQ2: } Weighing advertising performance and additional computation cost, is TFMS a cost-effective solution to solve the truncation problem?

{\bf RQ3: } What are the actual gains for the advertising system after adopting TFMS solution?

\subsection{Offline Simulation}

\begin{table}[htbp]
\caption{Truncation statistics for different targeting types}\label{tb:online-adset}
\begin{tabular}{lcc}
	\toprule
  \multirow{2}{2cm}{Targeting Type} & \multicolumn{2}{c}{Truncation Percentage}\\ 
   & Ad & User \\\midrule
   Retargeting & 43\% & 60\% \\
   Keywords Targeting & 9\% & 85\% \\
   Demographic Targeting & 36\% & 90\%\\
  \bottomrule
\end{tabular}
\end{table}

\begin{table*}[htbp]
\caption{Simulation results for truncation-free matching in online systems using real-world user and ad data}\label{tb:offline}
\begin{tabular}{lllllllll}
	\toprule
 	 Truncation-Free Stage & Time & RPM & PPC & PCTR & \# User-Crowd & \# User-Ad  \\ \midrule
  	(\emph{user} -> \emph{crowd}) & +26\% & +5.2\% & +4.4\% & +0.7\% & +57\% & +69\% \\
  	(\emph{crowd} -> \emph{ad}) & +17\% & +2.5\% & +1.8\% & +0.6\% & - & +112\% \\
  	both stages & +51\% & +5.6\% & +4.5\% & +1.1\% & +57\% & +259\% \\
  	\bottomrule
	\end{tabular}
\end{table*}

In Table~\ref{tb:trunc-stat}, we have already given some data about the percentage of truncated (\emph{user} -> \emph{crowd}) and  (\emph{crowd} -> \emph{ad}) pairs in our advertising system. We can see that on average 36\% of  (\emph{user} -> \emph{crowd}) pairs and 21\% of (\emph{crowd} -> \emph{ad}) pairs are truncated. Getting into other dimensions, Table~\ref{tb:online-adset} gives the online truncation statistics w.r.t. different targeting types. As the bridge between advertisers and users, the three typical types of targeting all suffer from severe truncation from the data, which brings obstacles for advertisers to effectively deliver their ads to the target users. To further quantify the influence of truncation about advertising performance, we conduct offline simulation experiments. 

The simulation is performed on a set of random-sampled display advertising traffic at Alibaba of \emph{Item Ads} scenario in a single day, which contains nearly one million real users. We replay these user traffic with the same setting as online systems in an offline simulation environment using real-world data. Three kind of truncation-free strategies are compared with base truncated two-stage matching: truncation-free matching in (\emph{user} -> \emph{crowd}) stage, truncation-free matching in (\emph{crowd} -> \emph{ad}) stage, and truncation-free matching in both stages.

Before giving the results, we firstly introduce the metrics that we use to estimate the performance. Revenue Per Mille, or RPM, is a commonly used metric in online advertising. It is the average advertising earning of every 1,000 ad impressions. In our simulation, the RPM is estimated by the predict click-through-rate (PCTR) times advertiser's cost per click (PPC). We also evaluate the response time and the volume of user-crowd and user-ad pairs calculated in matching system for each different strategy.

Simulation results are given in Table \ref{tb:offline}. We find that all of these three truncation-free strategies yield better RPM results. For the truncation-free matching in both stage strategy, the RPM metric increases 5.6\% compared to traditional truncated two-stage matching, which is significant enough in our application. Besides, the valid user-ad pairs number increases 259\%, which means that many advertisers gain additional chances to bid for their wanted user traffic. However, from the simulation, we find that the truncation-free matching strategy also causes 51\% of processing time increase in matching stage, which is unacceptable for online system.

\subsection{Computation Efficiency of TFMS}

Using near-line matching calculation, TFMS achieves truncation-free matching while brings little online latency and computation cost increase. However, to implement truncation-free matching, there are also other solutions, such as increasing the degree of online matching's parallelism along with more computation resource. To verify that whether the proposed TFMS is a cost-effective solution, we conduct quantitative comparison based on the aforementioned offline simulation experiments and the results are listed in Table~\ref{tb:offline-cost}. 

In the comparison, we use the number of user-ad pairs that the system needs to process to represent the computation cost of different solutions. Regarding that the traditional truncated two-stage matching's computation cost is $1$, TFMS's computation cost is no more than $0.86$ taking both the fully update and delta update into account~(it is hard for us to give accurate computation cost for delta update, however, in our real system, delta update takes less computation cost than fully update).
Correspondingly, the online truncation-free solution (denoted as `Online Parallelization' in the table) takes a computation cost of $3.59$, which is obviously larger than TFMS. Note that the fully update computation cost of TFMS is $0.43$, since that each daily active user will trigger $8.2$ times of advertising requests on average, while only needs once fully update in TFMS. With this strategy, TFMS is much more cost-effective compared with directly perform online truncation-free matching.

\begin{table}[htbp]
\caption{Truncation-free cost estimation based on offline simulation}\label{tb:offline-cost}
\begin{tabular}{llll}
	\toprule
  Method &  \# User-ad & Relative Scale \\\midrule
  Base & $\overline{\text{card}(\mathbf O_1(u))} * \#_{request}$ & 1\\
  Online Parallelization & $\overline{\text{card}(\mathbf O(u))} * \#_{request}$ & 3.59\\
   TFMS Fully Update & $\overline{\text{card}(\mathbf O(u))} * \#_{user}$ & 0.43\\
   TFMS Delta Update & Estimation value & <0.43\\
  \bottomrule
\end{tabular}
\end{table}

\subsection{Practical Online Results}

TFMS has been fully deployed in display advertising at Alibaba system since 2019, which brings great value for both the platform and advertisers. In our application, TFMS is deployed in two main display advertising scenarios, \emph{Banner Ads} and \emph{Item Ads}. The early deployment of TMFS is performed in an incremental way, i.e., reserving the traditional truncated two-stage matching pipeline while adding the additional TFMS pipeline at the same time, for the smoothness of advertising performance. The fully substitutional deployment of TFMS is carried out on schedule step by step. All targeting types except the automatic targeting are built in TFMS.

\begin{table}[htbp]
\caption{Performance on platform revenues in two ad scenarios in which TFMS is implemented}\label{tb:online-platform}
\begin{tabular}{llllllll}
	\toprule
  Scenario & Time & RPM & CTR & PPC \\\midrule
  Banner Ads & +1\% & +9.4\% & -0.4\% & +9.9\% & \\
   Item Ads  & +2\% & +5.1\% & -0.9\% & +5.9\% & \\
  \bottomrule
\end{tabular}
\end{table}  

\begin{table}[htbp]
\caption{Performance on advertiser winning impressions in different targeting types}\label{tb:online-advertiser}
\begin{tabular}{llllllll}
	\toprule
  \multirow{2}{2cm}{Targeting Type} & \multicolumn{2}{c}{Winning Impressions}\\ 
   & Banner Ads & Item Ads \\\midrule
   Retargeting & +12\% & +26\%\\
   Keywords Targeting & +23\% & +39\%\\
   Demographic Targeting & +112\% & +136\%\\\midrule
   All types & +51\% & +39\%\\
  \bottomrule
\end{tabular}
\end{table}

From both the advertising platform's view and the advertiser's view, the deployment of TFMS brings additional values, and the practical online results are listed in Table~\ref{tb:online-platform} and Table~\ref{tb:online-advertiser}. Firstly, we can observe that TFMS brings 9.4\% and 5.1\% RPM increase for \emph{Banner Ads} and \emph{Item Ads} respectively, which is a significant improvement considering the volume of advertising revenue at Alibaba. Secondly, as important tools for advertisers to select target audience, the main target types win more impressions after adopting truncation-free matching. For example, the total impression number of retargeting, keywords targeting and demographic targeting in \emph{Banner Ads} increases 51\% relatively, which brings more flexible biding experience for advertisers. 

\section{Conclusion}\label{sec:conclusion}

In this paper, we proposed a novel matching design TFMS to handle the truncation problem for matching module in display advertising at Alibaba. Online A/B test shows the efficiency of TFMS in both platform revenue and advertiser experience. TFMS not only gives a way to solve truncation problem, but more importantly, gives a novel possibility to explore matching methods in a way without online latency limitation. It's commonly believed that in online industrial systems, matching module cannot be too complicated due to limiting computing power. We believe that TFMS shows a possible way to handle more advanced models in ad retrieval, which is significant for many matching systems in display advertising.

\bibliographystyle{ACM-Reference-Format}
\bibliography{main}

\end{document}